\documentclass[galaxies,review,accept,pdftex,oneauthor]{Definitions/mdpi} 

\newcommand\aap{Astron. Astrophys.}                
\newcommand\apj{Astrophys. J.}                 
\newcommand\apjl{Astrophys. J. Lett.}                
\newcommand\araa{Annu. Rev. Astron. Astrophys.}             
\newcommand\mnras{Mon. Not. R. Astron. Soc.}             
\newcommand\nar{New~Astron.~Rev.}    
\newcommand\nat{Nature}              
\newcommand\prl{Phys. Rev.~Lett.}    
\newcommand\pasj{Publ. Astron. Soc. Jpn.}               
\newcommand\ssr{Space Sci. Rev.}     
\newcommand\src{Cyg~X-1}

\newcommand{\chandra}{\textit{{Chandra}}}
\newcommand{\nustar}{\textit{{NuSTAR}}}
\newcommand{\suzaku}{{{\it Suzaku}}}

\firstpage{1} 
\pubvolume{1}
\issuenum{1}
\articlenumber{0}
\pubyear{2024}
\copyrightyear{2024}
\datereceived{15 September 2024} 
\daterevised{9 November 2024} 
\dateaccepted{12 November 2024} 
\datepublished{ } 
\hreflink{https://doi.org/} 

\Title{Fifty 
	Years After the Discovery of the First Stellar-Mass Black Hole: A Review of Cyg X-1}

\TitleCitation{Fifty Years After the Discovery of the First Stellar-Mass Black Hole: A Review of Cyg X-1}

\Author{{Jiachen Jiang} 
 $^{1,2}$\orcidA{}}

\AuthorNames{Jiachen Jiang}

\AuthorCitation{{Jiang, J}}

\address{%
$^{1}$ \quad Department of Physics, University of Warwick, Gibbet Hill Road, Coventry CV4 7AL, UK; Jiachen.Jiang@warwick.ac.uk\\
$^{2}$ \quad {Institute of Astronomy, Madingley Road, Cambridge CB3 0HA, UK}}

\abstract{Around 50 years ago, the famous bet between Stephen Hawking and Kip Thorne on whether \src\ hosts a stellar-mass black hole became a well-known story in the history of black hole science. Today, \src\ is widely recognised as hosting a stellar-mass black hole with a mass of approximately 20 solar masses. With the advancement of X-ray telescopes, \src\ has become a prime laboratory for studies in stellar evolution, accretion physics, and high-energy plasma physics. In this review, we explore the latest results from X-ray observations of \src, focusing on its implications for black hole spin, its role in stellar evolution, the geometry of the innermost accretion regions, and the plasma physics insights derived from its X-ray emissions. This review primarily focuses on \src; however, the underlying physics applies to other black hole X-ray binaries and, to some extent, to AGNs.}


\begin{document}

\section{\src: The First-Ever Discovered Stellar-Mass Black~Hole}

\begin{figure}[H]
    \includegraphics[width=\textwidth]{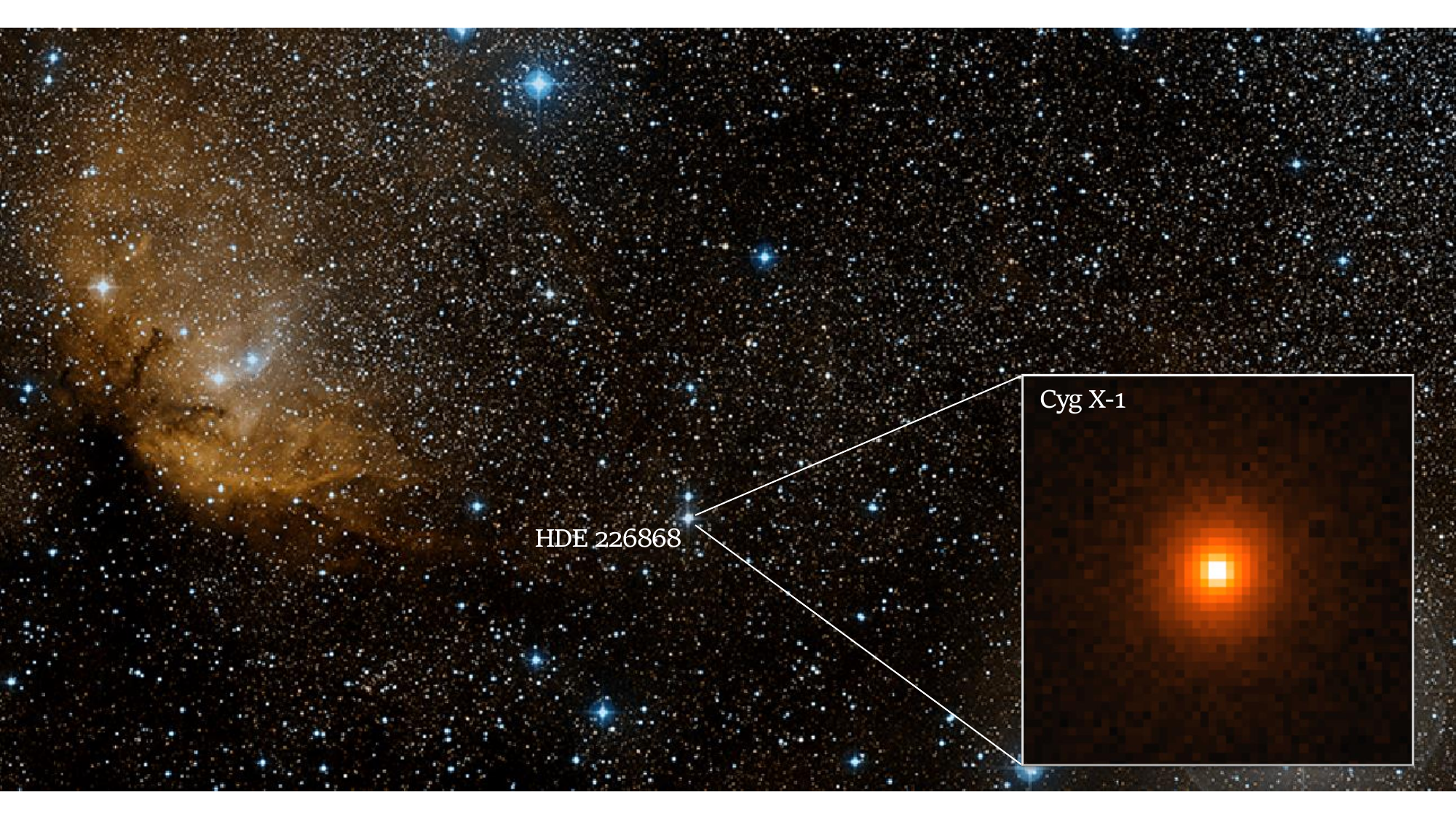}
    \caption{The ESO Digital Sky Survey optical image of the sky area of 60 by 30 arcmins around \src. The~blue dots in this image show the luminous young blue stars. One of them, the~supergiant star HDE 226868, has an invisible BH companion. The~accretion process from the star to the BH radiates significant X-ray emission. The~zoom-in image shows the \textit{{ROSAT}} image of the corresponding X-ray source \src. The~cloud to the left of the image is Sh2-101,  a~nearby H \textsc{ii} region emission nebula located in the constellation~Cygnus.}
    \label{pic_wfi}
\end{figure}

\src\ is one of the brightest persistent X-ray sources in the sky, with~an X-ray flux ranging from 0.2–2 Crab (see Figure~\ref{pic_wfi} for an image of \src). It was the first system widely believed to host a stellar-mass black hole (BH) after the work of \citet{bolton72,webster72}. The~discovery of \src\ as an X-ray source dates back even further, to~1964 \citep{bowyer65}. The~famous bet between Stephen Hawking and Kip Thorne in 1974 on the existence of a BH in \src\ \citep{hawking98} has since become a celebrated story in the scientific history of BHs. Fifty years on, advanced X-ray telescopes have enabled numerous discoveries across various aspects of X-ray binaries like \src. In~this review, I summarise some of the latest findings, focusing on X-ray~observations. 

The X-ray emission from \src\ primarily originates from the accretion of powerful stellar winds from its super-giant companion star \citep{orosz11}. A~parallax study by \citet{millerjones21} has precisely measured the distance to \src\ at $2.2\pm0.2$ kpc, providing a well-constrained BH mass of $21\pm2$\,$M_{\odot}$ and a companion star mass of $40$\,$M_{\odot}$. This massive companion star places \src\ among the high-mass X-ray binaries (HMXBs). Currently, there are only a small number of known BH HMXBs, as~listed in Table\,\ref{tab_hmxb}, among~which the nature of the compact object in MWC~656 is still debated \citep{janssens23}. The~rarity of BH HMXBs is largely due to selection effects: donor stars in low-mass X-ray binaries (LMXBs) typically have ages of $\geq$$10^9$ years, whereas those in HMXBs are much younger, with~ages of $\leq$$10^6$ years. Consequently, the~likelihood of observing a BH LMXB is higher than finding a BH HMXB. \src\ stands out as a unique example with the most massive BH and the second most massive donor star, following M33~X--7.

This review is structured around three major topics. First, I explore the role of \src\ in stellar evolution, focusing on its near-maximal BH spin, as~measured through relativistic disk reflection and thermal continuum emission. I discuss the implications of these measurements for theories of stellar evolution. Second, I focus on accretion physics, emphasising the innermost accretion geometry, particularly in light of recent X-ray polarisation measurements. Lastly, I cover plasma physics, focusing on high-energy pair plasma inferred from hard X-ray observations of the corona and~photo-ionised plasma inferred from soft X-ray observations of the accretion disk in \src. Although~this review centres on \src, many of the models discussed extend to other X-ray binaries or even active galactic nuclei (AGNs).

\unskip
\begin{table}[H]
        \caption{Known BH HMXBs in the local group. Values are from \citet{vandenheuvel18,millerjones21}. Note that the nature of the compact object in MWC~656 is still debated (e.g., \cite{janssens23}).}
        \newcolumntype{C}{>{\centering\arraybackslash}X}
    \begin{tabularx}{\textwidth}{CCCC}
    \toprule
       \textbf{Name}  &  \boldmath{$P$} \textbf{(Days)} & \boldmath{$M_{2}$} \boldmath{$(M_{\odot})$} & \boldmath{$M_{\rm BH}$} \boldmath{$(M_{\odot})$} \\
       \midrule
       \src  & 5.6 & $40^{+8}_{-7}$ & $21\pm2$ \\
       LMC~X-1 & 3.9 & $31.8\pm3.5$ & $10.9\pm1.4$ \\
       LMC~X-3 & 1.7 & $3.6\pm0.6$ & $7\pm0.6$ \\
       MWC 656 & $\approx60$ & $\approx13$ & $4.7\pm0.9$\\
       M33 X-7 & 3.45 & $70\pm7$ & $15.6\pm1.5$ \\
       \bottomrule
    \end{tabularx}
    \label{tab_hmxb}
\end{table}
\section{\src\ as the Laboratory for Stellar~Physics}

\subsection{Measure the Spin of the Black Hole in \src} \label{spincygx1}

The no-hair theorem posits that all stationary BH solutions can be completely characterised by only three independent parameters: mass ($M$), angular momentum ($J$), and negligible electric charge in an astrophysical expectation. The~angular momentum or spin ($J/M^{2}$) of a BH is particularly significant as it encodes information about the BH's formation history (e.g., by~mergers or accretion). Spin may theoretically play a critical role in powering jets that allow BHs to influence their surroundings. Most notably, high spin can produce extreme relativistic effects, such as frame-dragging, giving us a unique opportunity to test gravitational theories in the strong-gravity regime. The further implications of BH spin measurements can be found in \citet{reynolds19}, which I will not expand upon~here.

There are two established methods for measuring the spin of a BH---the relativistic reflection \citep{fabian89, reynolds14} and continuum fitting (CF, \citep{zhang97, mcclintock14}) methods. Both of these techniques work in a similar manner: by measuring the light from the innermost regions of the accretion disk, we attempt to determine where it terminates outside the BH and,~from that, infer the spin using the simple correlation between the innermost stable circular orbit and BH spin. The~CF method measures the temperature and luminosity of the accretion disk spectrum, which is hotter and brighter near the BH. The~reflection method instead measures how the fluorescent iron line in the X-ray spectrum, produced by X-ray reflection from the inner accretion disk, is distorted by the spacetime close to the BH, thereby inferring the inner~radius. 

\src\ was the first stellar-mass BH to be confirmed, and therefore, it has been extensively studied in the context of BH spin measurements. The~spin of \src\ has been measured to be higher than 0.95 using the continuum-fitting (CF) \mbox{method~\cite{gou14,kawano17,zhao21,kushwaha21}}. Relativistic disk reflection modelling has yielded spin values higher than 0.9 \citep{fabian12,tomsick14,parker15,duro16,walton16}. Although~alternative coronal geometries could introduce systematic uncertainties, the~spin value remains high. For~instance, \citet{krawczynski22} considered a cone-shaped coronal region illuminating the accretion disk and still found a disk spin of \mbox{at least~0.86}.

\subsection{The Missing Links Between \src\ and Gravitational Wave~Binaries}

The donor stars in HMXBs were believed to be massive enough to form a compact object, e.g.,~a neutron star (NS) or a BH, making them prime candidates for the progenitors of compact object binaries, as in the gravitational wave (GW) sources (e.g., \cite{ligo16}). However, the~inconsistency between their BH spin distributions was soon noticed by the community (e.g., \cite{belczynski20,fishbach22}). Accretion after BH formation was proposed to explain the high BH spins in BH HMXBs. However, since the accretion process cannot occur for longer than the lifetime of the donor \citep{millerjones21}, the~observed close-to-maximum spin of \src\ would require a persistent super Eddington accretion process \citep{moreno11}, but~\src\ is only observed to accrete with $2\times10^{-7}M_{\odot}$ yr$^{-1}$ now. The~actual accretion time may be even shorter than the age of the donor star, making the spin-up-due-to-accretion scenario more challenging \citep{russell07}. 

{The spin discrepancy between \src\ and GW binary BHs motivated a recent development in a modified relativistic disk model, including a `warm' corona for the X-ray data of \src\ that may lead to a lower inferred BH spin \citep{zdziarski24},} but such a development does not show significant statistical improvement in data fitting compared to the previous models \mbox{(e.g., \cite{gou14})}. The~condition of a `warm' corona in \src, similar to the ones proposed for supermassive BH accretion disks in AGNs with dramatically different BH disk temperatures and luminosities \citep{petrucci20},  has yet to be examined \citep{gonkiewicz23}.


\subsection{The Origin of the High BH Spin in \src}

In this review, I would like to highlight the large uncertainty on whether the known population of BH HMXBs can contribute significantly to the observed GW binary BHs and whether they make comparable~systems. 

Take \src\ as an example. \citet{neijssel21} conducted a population synthesis test based on the donor star and BH mass measurements in \citet{millerjones21}. They found that \src\ will have only a $7\%$ chance of forming a BH-NS binary and remaining bound after the NS natal kick. After~revising mass transfer models, \citet{neijssel21} found the probability of \src\ forming a binary BH within a Hubble time is 4\%. Similar conclusions were achieved for the other BH HMXB LMC~X-1 \citep{belczynski12}. 

The close-to-maximum spin of \src\ inferred by the X-ray data likely results from the angular momentum of the progenitor star's core instead of purely due to accretion. For~example, \citet{galleogos22} found that high-spin BH HMXBs can form through main sequence mass transfer from the BH progenitor to the secondary star (Case A mass transfer). The~core of the progenitor is tidally locked and, hence, rapidly rotating, which can produce high BH spins \citep{qin19}. Evidence of such a stellar evolution pathway includes the observed enhanced nitrogen abundances in the donor star of \src\ \citep{shimanskii12}.  For~the Case-A mass transfer model, mass is transferred from deep layers of the progenitor star, reprocessed from the CNO cycle (for carbon–nitrogen–oxygen) and is thus nitrogen rich. This evolution pathway is also supported by the low orbital eccentricity of \src\ \citep{millerjones21}. 

Alternatively, it was also suggested that slow ejecta from a failed supernova that formed the BH could interact with the companion and be torqued, increasing their angular momentum before falling back onto the newly formed BH \citep{batta17}. Hydrodynamical simulations in \citet{schroder18} show that such a spin-up process is consistent with some of the GW binary BHs at the higher end of the spin distribution, but~whether it is sufficient to spin up to close-to-maximum spin as in \src\ remains a~question.

Furthermore, the~significant BH mass difference between GW binary BHs and observed BH HMXBs also indicates their fundamental difference. The~primary BH mass distribution compiled from GW observations peaks around $35M_{\odot}$ \citep{abbott22b}. No observed BH HMXBs are close to this mass limit (see Table~\ref{tab_hmxb}).

Suppose that \src\ may not represent the best candidate progenitor for GW binary BHs due to significantly different BH masses and spins. Can we extend such a conclusion to other HMXBs that have yet to be detected? \citet{liotine23} used population synthesis calculations and applied realistic X-ray and GW observational limits of \chandra\ 
 and advanced LIGO. They found, due to observational selection effects, only a 3\% probability of a detectable HMXB host BH of >$35\,M_{\odot}$. The~probability of finding a HMXB forming a binary BH system as the GW sources within a Hubble time is only 0.6\%. Therefore, unsurprisingly, the~currently observed BH HMXBs, like \src, do not resemble the features of GW binary~BHs.  

\subsection{Black Holes in the Mass~Gap}

The findings of short-period, high-BH mass GW binaries are intriguing as they surpass the masses predicted by most stellar evolution models and exceed the known stellar-origin BHs in our galaxy: stars of an initial mass of higher than $30\,M_{\odot}$, as~the donor star of \src, are expected to lose significant mass and only produce BHs of <$20\,M_{\odot}$ \citep{vink08,sukhbold16}. 

{Isolated stellar-mass BHs with masses of a few tens of solar masses, formed through typical stellar evolution, are challenging to detect. Gravitational microlensing techniques can be employed to identify these isolated BHs (e.g., \cite{wyrzykowski20, lam22, lam23}), and~some candidates have been identified with masses in the range of~10--100\,$M_{\odot}$ \citep{kaczmarek24}.}

{A prevailing hypothesis proposes that these high-mass stellar-mass BHs bound in a binary system, such as those detected as GW sources, are remnants of massive, metal-poor stars. Low metallicity leads to notably less mass loss due to metallicity-dependent winds during the star’s lifetime \citep{vink08}. Additionally, the~reduced metallicity decreases the radius of the progenitor, lowering the likelihood of mergers during the common-envelope phase \citep{hurley00, belczynski07}. Finally, the~higher BH mass resulting from low-metallicity progenitors corresponds to weaker natal kicks at formation, which increases the chances of preserving the binary system as a bound pair \citep{belczynski10}.} This scenario is particularly supported by the new discovery of a $32.7\pm0.8\,M_{\odot}$ BH in a binary system with a period \mbox{of 11.6~years}, identified through \textit{{Gaia}} astrometric data \citep{gaia24}. The~broad-band photometric and spectroscopic data also confirm the existence of a very metal-poor, old companion star at a distance of 590 pc, suggesting that metal-poor environments may play a crucial role in the formation of such massive BHs \citep{gaia24}.

\section{\src\ as the Laboratory for Accretion~Physics}


\subsection{The Persistently High Radiative~Efficiency}

\begin{figure}[H]
    \includegraphics[width=\textwidth]{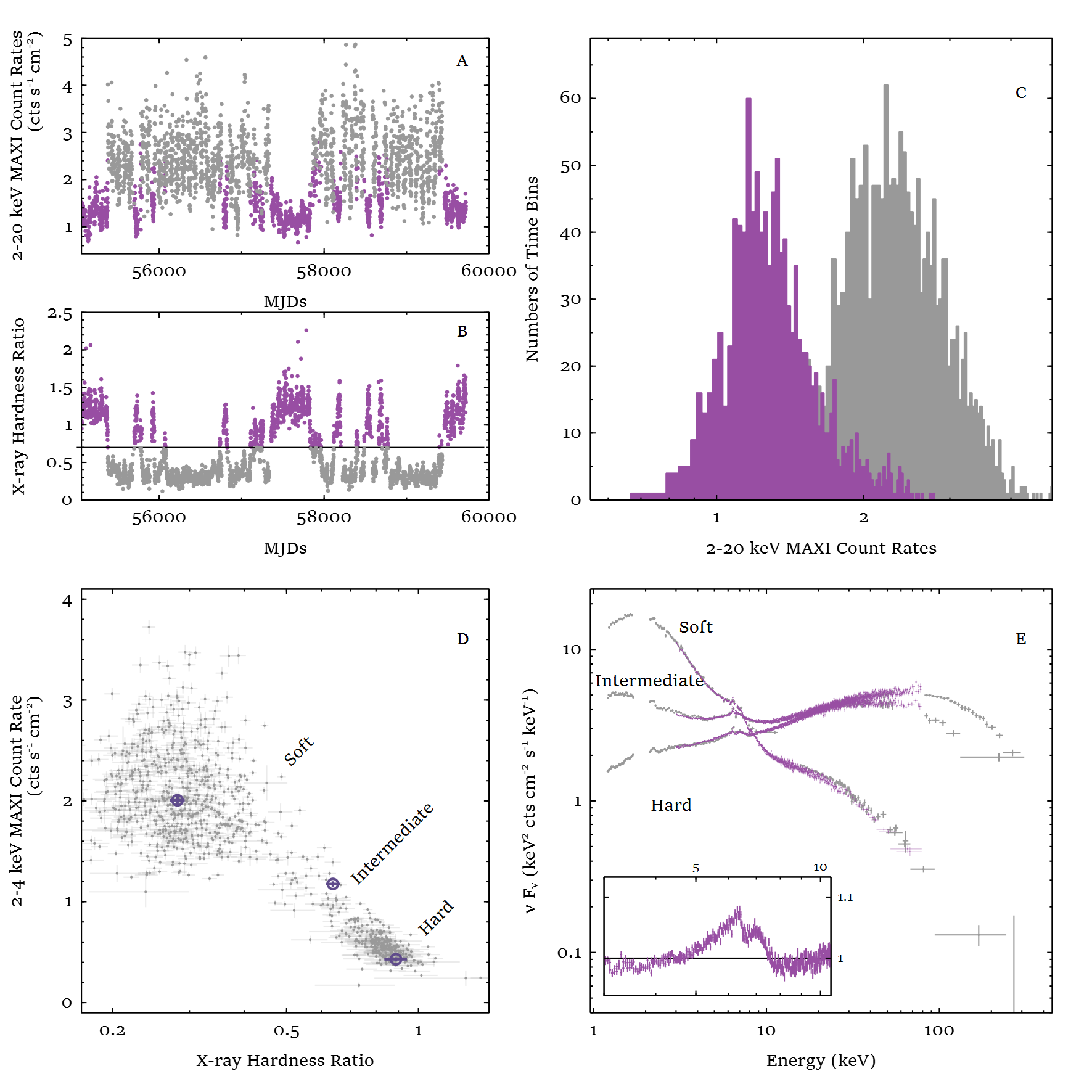}
    \caption{(\textbf{A},\textbf{B}): \textit{{MAXI}} {2$-$20 keV} 
 daily light curve and hardness ratio of \src. (\textbf{C}): The distribution of \textit{{MAXI}} X-ray count rates of \src\ shows a double-peaked log-normal distribution. \mbox{The~purple (grey)} distribution corresponds to the hard (soft) state. (\textbf{D}): \textit{{MAXI}} hardness-intensity diagram. (\textbf{E}): Quasi-simultaneous \nustar\ and \suzaku\ spectra of \src\ in the soft, intermediate and hard states. Their corresponding hardness ratio was marked by purple in Panel D. The~inset window shows the zoom-in of the hard state spectrum of \src, demonstrating significant evidence of relativistic disk Fe K emission lines. Detailed analyses of these spectra can be found in \citet{tomsick14,parker15,tomsick18,walton16}.}
    \label{pic_hid}
\end{figure}

Unlike many other BH transients, \src\ has never reached the quiescent state since its discovery in the X-ray band. Figure\,\ref{pic_hid} shows the X-ray brightness of \src\ as measured by \textit{{MAXI}}. The~\textit{{MAXI}} X-ray count rate alternates between high and low states, both of which exhibit significant variability. This behaviour results in a double-peaked log-normal distribution, as~shown in Panel C. The~two peaks correspond to different spectral states: the soft state, marked in grey, where the \textit{{MAXI}} count rate is high and the hardness ratio is low; and the hard state, marked in purple, where the \textit{{MAXI}} count rate is low and the hardness ratio is high. As~the X-ray flux decreases, the~hardness ratio correspondingly increases. This creates two distinct clusters in the hardness-intensity diagram (HID) shown in Panel D of the same figure. During~transitions from the hard to soft state, \src\ briefly passes through an intermediate state. Different X-ray spectral states were realised soon after the X-ray data of \src\ became available (e.g., \cite{tananbaum72}).

It is important to note that, due to historical reasons, the~X-ray HIDs of transients are typically presented using X-ray count rates. However, these count rates do not accurately reflect the true X-ray flux, as~the photon-collecting area of the telescope varies across different energy bands. Despite the fact that the \textit{{MAXI}} count rate appears lower in the hard state, the~actual X-ray luminosity is not significantly reduced. This is because the spectrum is harder, meaning each photon carries more energy. For~a more detailed discussion on instrument-dependent state transitions, see \citet{grinberg13}. 

Panel E illustrates the quasi-simultaneous \suzaku\ and \nustar\ spectra of \src\ during the soft, intermediate, and~hard states, corresponding to the three highlighted positions in the HID shown in Panel D. The~Eddington ratios of \src\ during these observations were measured to be around 1\%, with~values ranging from 0.62\% to 1.72\% (e.g., \cite{tomsick18}), corresponding to, \mbox{as~\citet{konig24}} demonstrates, the~lower end of the classical q-shaped HID typically observed in LMXBs. The~persistently high accretion rate is likely a consequence of the BH being embedded in the stellar winds of its massive donor star. Similar persistent X-ray emission is observed in other BH HMXBs, such as LMC~X-1 and LMC~X-3.

{\src\ has not been observed in a quiescent state, unlike many BH transients in LMXBs \citep{kong02}, where low accretion rates lead to the formation of a radiatively inefficient, optically thin flow due to the importance of viscous heating advection (e.g., \cite{yuan14}). In~contrast, the~presence of significant disk thermal and reflected emissions in both the soft and hard states of \src\ suggests the existence of an optically thick accretion disk extending close to the innermost stable circular orbit (e.g., \cite{gou14,parker15}).}

{Extending the discussion of accretion flow structures from \src\ to a broader range of BH XRBs, the~precise Eddington ratio at which the transition occurs---from an optically thin, radiatively inefficient, advection-dominated accretion flow to an optically thick, radiatively efficient, geometrically thin accretion disk, as~observed in \src---is still debated within the observational community, though~it is generally thought to lie between 0.1\% and 10\% of the Eddington luminosity. For~example, some studies, such \mbox{as \citet{plant15}}, have identified a significantly truncated, optically thick accretion disk at a few tens of gravitational radii in GX 339$-$4 at an Eddington ratio of a few percent. In~contrast, other groups report an optically thick disk extending to the innermost stable circular orbit for the same object at a similar luminosity \citep{garcia15}. The~formation of an optically thick accretion disk may depend on the specific environment of each object. However, \src\ is distinctive in that it remains around 1\% of the Eddington luminosity, serving as a unique laboratory for accretion studies---a topic I will expand on in a later section.}

{Future high-resolution X-ray spectroscopy, such as that from \textit{{XRISM}} equipped with microcalorimeters, will be crucial in revealing the nature of accretion disks through observations in the iron emission band. \textit{{XRISM}} has already demonstrated remarkable spectral measurement capabilities, disentangling multiple iron line emission components when used jointly with CCD-resolution data (e.g., \cite{xrism24a, xrism24b}). As~of this writing, analysis of the \textit{{XRISM}} observations of \src\ is ongoing.} 


\subsection{The Accretion Geometry of \src}

\textbf{{Coronal geometry:} 
} the geometry of the X-ray Comptonisation region---where disk seed photons are Compton-upscattered by an electron cloud with an optical depth of $\tau \approx$ 0.5--2 \citep{thorne75,sunyaev79}, often referred to as the corona---remains a topic of debate. Despite the uncertain geometry, the~size of the corona is known to be small in various accreting BH systems, including stellar-mass BH transients and AGNs, as~evidenced by reverberation and gravitational lensing studies \citep{fabian09,kara19,reis13,chartas16}. Additionally, the~steep emissivity profile of the accretion disk also suggests a compact corona \citep{wilkins12}. The~latter was investigated in detail for \src\ using the \suzaku\ data \citep{fabian12}. The~emissivity profile of the disk in \src\ is consistent with the model for a compact coronal region with a size of 5--7\,$r_{\rm g}$.

The geometry of the corona is closely linked to the question of its origin, making it essential to constrain the corona's geometry through observations. Historically, the~``lamppost'' geometry, in~which a point source is located along the BH's spin axis, has often been assumed when modelling the X-ray spectra and timing properties of accreting BH transients, including \src\ (e.g., \cite{parker15}). This approximation is based on theories suggesting a jet-based origin for the corona \citep{henri97,miniutti03,markoff05}. 

However, the~``lamppost'' model should not be taken too literally; it is merely a simplification representing a very compact coronal region near the BH. We often adopt this point-like geometry because it is computationally convenient. For~example, setting the initial condition along the spin axis in relativistic ray-tracing calculations is far simpler than modelling more complex geometries. Nonetheless, several alternative geometries have been proposed, such as a spherical or outflowing jet-shaped corona \citep{gonzalez17}, sandwich or wedge layers above the disk \citep{galeev79,stern95}, and~thick accretion flows within a truncated disk \citep{eardley75}.

{Surprisingly, shortly after the launch of \textit{{IXPE}}, \src\ was found to exhibit a higher-than-expected polarisation degree in the hard state. While a low polarisation degree of less than 1\% was anticipated from the corona, {observations in the 2--8\,keV band revealed a polarisation degree of around 4\%} {\citep{krawczynski22b}}.} 
Moreover, the~polarisation angle aligns with the AU-scale jet position angle of \src\ in the sky, as~well as with the pc-scale radio lobes~\citep{gallo05}. This suggests that the corona may be extended over the plane of the accretion disk, assuming the polarisation originates entirely from the corona rather than the~jet. 

However, a~horizontally extended coronal region does not fully explain the unexpectedly high polarisation degree. \citet{krawczynski22b} proposed that if the corona is a layer on the disk with the same inclination as the disk, then the inner disk must have a significantly higher inclination than the binary’s orbital plane \mbox{(e.g., around $30^{\circ}$~\citep{millerjones21})}. Previous disk reflection spectral models indeed suggested a possible misalignment of around $13^{\circ}$ \citep{tomsick14}.

Alternatively, if~the corona is outflowing at a relativistic velocity of $\geq$0.4\,$c$, the~angular distribution of seed photons would be significantly different in the plasma's rest frame. This would result in a higher polarisation degree for the outgoing scattered radiation without the need for a high inclination angle to explain the data \citep{poutanen23}. 

It is crucial to consider the implications of such high velocities within the corona. As~discussed in a later chapter, the~compact, hot corona is primarily composed of pair plasma. A~key question remains whether such a high velocity might exceed the equilibrium velocity of the pairs, the~exact value of which depends on the optical depth of the pair plasma and is estimated to be around 0.3--0.4 $c$ for the condition similar to \src\ \citep{beloborodov99}. Additionally, a~highly outflowing corona would result in a significantly lower disk reflection fraction compared to a static corona model \citep{dauser16,you21} because of the beaming effects. A~fully consistent disk reflection model for the spectral and polarisation data of \src\ that incorporates the consistent coronal geometry has yet to be~developed.

Last but not least, it is important to acknowledge two key points. First, the~geometry of the corona may be highly source-dependent. For~example, if~the corona is magnetised (e.g., \cite{liu02,miller00}) or regulated by radiation, both mechanisms depend on the specific model used or the target observed, making it challenging to draw broad conclusions from a small sample of sources. Second, due to the limitations of current observational methods, we are still in the phase of testing different geometries by making assumptions within models. Nature often behaves in unexpected ways, so the scientific community must remain open-minded. While testing specific geometries, it is crucial not to exclude other possibilities or the potential combination of simple geometries, e.g.,~a disk-shaped lamppost corona that corotates with the disk of \src. An~effort in this direction is being developed (Bambi, in~private communication).  The~size of the corona is not constrained by polarisation data but can well be by timing and spectral data. A~joint analysis of timing and spectral and polarisation data has yet to be~complete.

\textbf{{Links to the jet:}} the explanations above do not entirely rule out a jet origin for the 2--8\,keV polarisation observed in \src. \citet{dexter23} proposed that a weakly polarised seed emission, such as that expected from a standard lamppost corona or the disk, could be Compton-upscattered by cold electrons\endnote{In this model, ``cold electrons'' refers to those with  lower Lorentz factors than the bulk Lorentz factor of the jet.} in the jet, which is the so-called bulk Comptonisation model.  A~highly polarised scattered emission can result from this interaction because relativistically moving electrons have a higher scattering cross-section for anisotropic radiation beamed in the direction of their motion \citep{begelman87}. 

To understand \src\ in the hard state, the~community ought to explore synergistic opportunities. Moving on to other wavelengths, the~high polarisation degree of \src\ in the soft Gamma-ray band, e.g.,~$>$20\% \citep{rodriguez15},  is likely due to the synchrotron emission from the jet, which is prominent in the hard state, as~indicated by its strong radio emission, and~typically quenched in the soft state. The~jet's soft gamma-ray emission overlaps with the X-ray Comptonisation component at lower energies and is associated with the radio jet. These features align with the multi-component emission processes predicted by compact jet models, such as those proposed by \citet{markoff05}. A~recent comprehensive study of the multi-wavelength polarisation properties of Cyg X-1 \mbox{by~\citet{russell13}} demonstrated that the spectrum and polarisation are consistent with a compact jet dominating the radio to the infrared range and contributing to the MeV~tail. 

Intriguingly, \textit{{INTEGRAL}} observations found a polarisation angle of $40^{\circ}$\linebreak \mbox{above 230\,keV~\citep{jourdain12,rodriguez15}, }which is $60^{\circ}$ different from the angles observed in optical, infrared,~\linebreak\mbox{2--8\,keV} bands as well as the radio jet orientation \citep{fender06,russell13}. \textit{{AstroSAT}} found an almost $90^{\circ}$ difference between the 100--380\,keV energy band and radio jet orientation \citep{chattopadhyay24}. A similar deviation between radio jet orientation and soft gamma-ray polarisation angles was also discovered in the hard state observations of other BH XRBs (e.g., \cite{bouchet24}). This discrepancy suggests possibly complex magnetic field structures or multiple electron populations within the jet. To~fully understand the hard state of \src, it is essential to develop consistent multi-wavelength models that account for the jet’s~contribution.

\textbf{{Disk geometry:}}  \citet{steiner24} reports that \textit{IXPE} measurements of \src\ in the soft state show a polarisation angle consistent with the radio jet orientation. The polarisation degree is two percent lower than in the hard state. \textcolor{black}{Similar conclusions were found in \citet{jana24}.} Similar to the hard state, the polarisation degree in the 2--8\,keV band increases with energy. Comparable results have been observed in the soft states of other LMXBs \citep{ratheesh24, rodriguez23, podgorny23, marra24, svoboda24}. 

The very high polarisation degree observed in the soft state, dominated by disk thermal emission, suggests that the classical standard disk model, which assumes a semi-infinite, free-electron scattering atmosphere \citep{chandrasekhar60}, may no longer be applicable. Additionally, general relativity predicts a decrease in polarisation degree with increasing energy, which contrasts with observations of \src\ and other objects, such as 4U~1630$-$47 \citep{rodriguez23}.

One possible solution to achieve such a high polarisation degree, proposed by \citet{ratheesh24}, involves a scattering atmosphere in 4U~1630$-$47 with an outflow velocity of $0.5$\,c—similar to the velocity assumed for the hard-state outflow\footnote{Notably, this outflow in the soft state involves particles, while the hard state involves a pair plasma outflow.}. In such a disk model, absorption within the outflowing atmosphere results in a high degree of polarisation parallel to the disk surface. While this model may not align with the observed polarisation angle for \src, it could apply to other objects with different polarisation orientations, such as 4U~1630$-$47.

An alternative method to achieve higher polarised flux is through reflection due to disk self-irradiation, which is more likely the case for \src\ \citep{steiner24}. This model has also been successful for other objects \citep[e.g.,][]{marra24}. In this scenario, harder X-rays from a hotter region of the accretion disk that is closer to the BH are more likely to be reflected back to the disk. This reflection naturally increases the polarisation degree with energy, consistent with the observations of \src. {It is important to note that the strength of returning radiation strongly depends on the BH spin. Only systems with extremely high BH spin, such as} \src, can generate significant polarisation through a self-irradiated disk. This is consistent with previous spectral measurements of the BH spin in \src\ (see Section \ref{spincygx1}). 

However, it is crucial to note that the polarisation degree of a self-irradiated disk depends on the disk geometry. The assumption of a razor-thin disk model may be valid for \src, given its consistently low luminosity (less than 10\% of its Eddington luminosity). This assumption may be less applicable to other BH transients \citep[e.g.,][]{straub11, fabian20}, where luminosity can commonly peak around the Eddington limit in the soft or ultra soft states.

\subsection{Fast X-Ray~Variability}

\textbf{{Variability:}} the study of variability on short timescales is often conducted in the Fourier frequency domain using the power spectral density (PSD) spectra. 

The hard state of \src\ exhibits broad-band noise that can be decomposed into multiple Lorentzian components \citep{nowak00, pottschmidt03, konig24}. Occasionally, some very high-frequency Lorentzian components are required to fit the PSD \citep{pottschmidt03, axelsson18}, although~they do not always persist consistently. These Lorentzian components remain at the same Fourier frequency, but~their strength varies with energy: harder X-rays show more variability than the softer X-rays. For~instance, \textit{{NICER}} measured the root-mean-square (RMS) variability, integrated over all Fourier frequencies, to~be 19\% in the 0.5--1\,keV band and a higher value of 30\% in the 5--8\,keV band \citep{konig24}.

Some of the extreme variability observed in the hard state is likely linked to the same processes responsible for the typical X-ray variability seen during smaller flares \citep{uttley05}. This suggests that the powerful flares in the low/hard state are part of the same variability mechanism driving the more frequent, lower-amplitude fluctuations. This view aligns well with the exponential model, which predicts a lognormal distribution of fluxes \citep{gierlinski03}. 

In the intermediate state, the~PSD of \src\ reveals a broad Lorentzian component, which predominantly influences the hard X-ray band above 3 keV. In~the soft X-ray band, where disk thermal emission dominates, the~RMS variability is significantly lower, although~the PSD retains a similar shape \citep{konig24}.

In the soft state, the~variability of \src\ exhibits typical red noise behaviour. Similar to the intermediate state, the~RMS variability increases from 7\% in the 0.5--1\,keV band \mbox{to 30\% in the 5--8\,keV} band \citep{grinberg14, zhou22}. The~overall RMS variability across the 0.5--10\,keV range is around 10\% \citep{cui97}, which is substantially higher than that of most LMXB transients in the soft state, where the RMS is typically below 1\%, such as in MAXI J1820 $+$ 070 \citep{li24}.

Several of the previous studies have demonstrated that the X-ray timing characteristics of other accreting compact objects exhibit a linear relationship between their RMS variability and average flux across various timescales (e.g., \cite{uttley01}). This relationship suggests non-stationarity in the light curves, where the level of instantaneous variance is not constant but instead scales with the longer-term flux averages. \citet{uttley05} highlight that these properties imply a lognormal distribution of instantaneous flare strengths. This can be explained if the observed variations are not the result of a superposition of independent shots but~rather due to accretion rate fluctuations occurring at all locations within the accretion flow. These fluctuations propagate inward, coupling perturbations at different radii. Interestingly, a~similar ``RMS–flux relation'' has been observed in other systems, such as narrow-line Seyfert galaxies \citep{gaskell04}, suggesting that this relation may universally constrain the fluctuating conditions of hot~plasmas.

\textbf{{Coherence and lags:}} \src\ exhibits complex behaviour in terms of coherence and time lags, which vary depending on the choice of energy bands and Fourier frequencies. I will not demonstrate the intricate details in this review. Readers may instead refer to the series of work in the past decades \citep{pottschmidt03,gleissner04,wilms06,grinberg14,konig24}. There are numerous uncertainties in their interpretations: for instance, while lags are well documented in the soft state, where the disk thermal emission dominates the X-ray band \citep{grinberg13, konig24}, the~origin of these lags remains~unclear.

In the hard and intermediate states, where RMS variability is higher and the coherence between soft and hard X-ray variability is greater than in the soft state, both hard and soft lags are detectable and much better understood than in the soft state. The~hard X-ray lag, relative to the soft X-rays, increases with energy, supporting propagating fluctuation models \citep{lyubarskii97}. While the emission might originate primarily from the innermost regions, the~variability likely arises throughout the entire accretion flow. In~particular, \citet{arevalo06} showed such a propagation model for \src\ in great~detail.


The X-ray spectra of \src\ across various accretion states exhibit substantial evidence of disk-reflected emission, with~the most prominent feature being relativistic broad Fe K emission lines spanning the 6–7 keV range. Interestingly, no clear iron line lag features as in LMXBs (e.g., \cite{kara19}) have been detected in archival observations of \src\ from \textit{{RXTE}}~\citep{mastroserio19}, \textit{{NICER}}~\citep{konig24}, or~Insight-\textit{{HXMT}}~\citep{zhou22}, likely due to the limited reflection strength in the iron emission band. However, significant soft lags have frequently been observed~\citep{konig24}, which have been interpreted as the light-crossing time between the hard X-rays from the corona and the soft X-ray photons from the disk, similar to findings in other systems \citep{uttley11,wang21}.

\section{\src\ as the Laboratory for High-Energy Plasma~Physics}
\unskip

\subsection{Optically Thin $10^{9}\,K$ Pair~Plasma}

Electron--positron plasma is of significant interest across various fields of physics. For~instance, pairs play a crucial role in the dynamics of gamma-ray burst expansion \mbox{(e.g., \cite{goodman86})} and contributed to the evolution of the early Universe (e.g., \cite{kolb90}). The~relevance of pairs has become especially prominent with the availability of high signal-to-noise hard X-ray data from \nustar\ \citep{fabian15}: in the hard X-ray band, accreting black holes are observed to radiate significantly above the electron rest mass energy \citep[511 keV,][]{guilbert83}. This high-energy emission above 511 keV is constrained by electron--positron pair production within the compact region of the corona \citep{svesson82}, establishing a hard upper limit on coronal temperature relative to the given luminosity. By~compiling \nustar\ measurements of electron temperatures ($kT_{\rm e}$) for a sample of AGN and a few BH X-ray \mbox{binaries,~\citet{fabian15}} found that most sources reside near this hard limit, suggesting pair-dominated coronae in these~systems.

Among sources with measurements of $kT_{\rm e}$, cool coronae are relatively rare in both AGN and XRBs (e.g., \cite{tazaki10, kara17, tortosa17, walton18b, xu18, jiang19b, buisson19, zhang22}). These low-$kT_{\rm e}$ sources exhibit a wide range of Eddington ratios, from~just a few percent \citep{tortosa17} to nearly Eddington levels \citep{kara17}. The~inclusion of pair production from non-thermal components within the plasma can account for the observed lower temperatures in these sources \citep{fabian17}. The~hybrid plasma model predicts that an increased non-thermal fraction in the corona will lead to a rise in hard X-ray flux and the appearance of the annihilation line, along with a decreasing high-energy cutoff in the spectrum \citep{coppi99}.

\begin{figure}[H]
    \includegraphics[width=\textwidth]{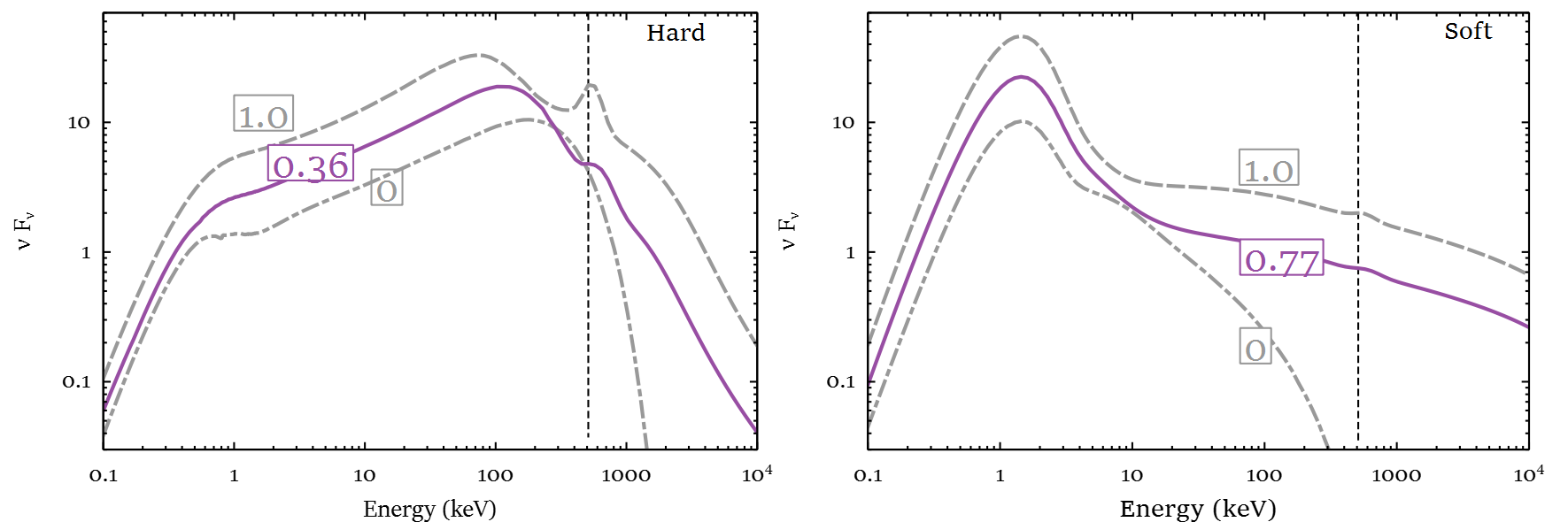}
    \caption{The best-fit hybrid pair plasma models for the coronal emission component of \src\ in the soft and hard states (in purple). The~models were calculated using \texttt{Eqpair} \citep{coppi99} based on the best-fit parameters in \citet{gierlinski99,nowak11}. The~grey models show the spectra for a different fraction of power supplied to the non-thermal distributed particles. Notice the significant increase in hard X-ray emission and the prominent annihilation line at 511\,keV when the fraction of non-thermal particles increases. When the non-thermal fraction decreases to zero, this model is consistent with the purely thermal Comptonisation model (e.g., \cite{titarchuk94}). Note that the annihilation line at 511\,keV is still present but only is broadened when the non-thermal fraction is~low.}
    \label{pic_plasma}
\end{figure}

To demonstrate the effect of a hybrid thermal and non-thermal electron energy distribution, I present the hybrid plasma models for both the hard and soft state spectra of \src, using best-fit parameters from \citet{nowak11} and \citet{gierlinski99} in Figure\,\ref{pic_plasma}. For~simplicity, I only show the disk thermal and coronal emission components. The~grey models illustrate the spectral variation when altering the fraction of power supplied to non-thermal particles. Notably, the~511 keV annihilation line is always present but experiences much greater Doppler broadening when the plasma is thermally dominated, allowing for a higher electron~temperature.

For Cyg X-1, explaining the observed hard X-ray excess emission of \src\ solely through pure thermal Comptonisation is challenging (e.g., \cite{titarchuk94}), a~point that has not been fully appreciated: hybrid plasma-based spectral modelling for \src\ has revealed evidence of a hard X-ray excess above 100 keV, as~observed by \textit{{Ginga}}, \textit{{OSSE}}, and~\textit{{COMPTEL}}~\citep{ling1987, poutanen98, gierlinski99}. Later, the~measurements of the electron distribution in the corona of Cyg X-1 were refined with data from GSO, the~well-type phoswich counter aboard \suzaku, which provided simultaneous hard X-ray data in complement to soft X-ray data from CCD instruments (e.g., \cite{nowak11, parker15}).

Conducting this type of analysis has proven challenging for several reasons. \linebreak\mbox{\citet{markoff05}} proposed that hard X-rays are dominated by inverse Compton scattering at the jet base, with~both disk and synchrotron photons serving as seed photons to explain the data. According to this model, the~corona at the jet base primarily contributes to hard X-rays, while synchrotron emission dominates from radio through soft X-rays, providing an adequate fit. \citet{nowak11} tested both the hybrid plasma model and jet models, finding that they work for the hard state up to 300 keV for \src. However, constraining these models with the currently available data remains difficult. The~future detection of an annihilation line in \src\ will be crucial for resolving these issues. In~the soft state, when the radio jet is quenched, hard X-ray emission persists, but~the flux is too low to obtain a high-quality spectrum for testing the hybrid plasma model. In~summary, high-signal-to-noise observations in the hard X-ray to soft gamma-ray range, such as those possible with \textit{{COSI}} \citep{tomsick23}, are needed to advance this~research.

\unskip

\subsection{Optically Thick $10^{5}\mbox{--}10^{7}\,K$ Plasma}


One of the most well-established Eddington-limited analytical accretion models is the standard thin disk model, developed in the 1970s \citep{shakura72} and still widely used today. In~this model, the~disk is characterised as a geometrically thin, optically thick accretion flow with multi-colour blackbody emission. This model is consistent with the optical and UV emission observed from accreting SMBHs in AGNs and the soft X-ray emission from BH XRBs. To~test the standard accretion disk model, one could examine several direct outcomes of this model, such as the geometry of the accretion disk (e.g., \cite{jiang22x}) or the estimation of the emission region based on reverberation studies (e.g., \cite{kammoun19}).

Spectrally, the~standard thin accretion disk model can also be tested using atomic spectra from the accretion disk, with~densities predicted by the model. Additionally, the~high temperatures and densities of BH accretion disks in strong-gravity environments make them challenging to study in laboratory settings or convincingly simulate computationally. Therefore, X-ray observations of BH accretion disks are unique and crucial for understanding high-energy plasma with extreme~properties.

Previous observational studies have explored the relationship between BH mass, mass accretion rates, and~the density of the accretion disk in AGNs and XRBs \citep{jiang19c, jiang19d}, which roughly agrees with the expectation in the standard thin disk model. This critical aspect has often been overlooked in the prior~research. 

Interestingly, an~attempt to fit a high-density model to the intermediate state spectra of \src\ suggested an electron density of approximately $10^{18}$ cm$^{-3}$ \citep{tomsick18}. Compared to a low-density model with an electron density fixed at $10^{15}$ cm$^{-3}$, the~high-density model\endnote{A high density usually refers to values significantly higher than $10^{15}$ cm$^{-3}$, which were previously assumed in disk models. For~stellar-mass BH accretion disks, high density would range between $10^{18\mbox{--}22}$ cm$^{-3}$, while for supermassive BH accretion disks, it usually falls between $10^{15\mbox{--}19}$ cm$^{-3}$.} yields a higher reflection strength and implies lower iron abundances. This approach helps address the systematic overestimation of disk iron abundance seen in previous models, though~it may not be applicable to all XRB spectra \citep{liu23}.

\begin{figure}[H]
    \includegraphics[width=\textwidth]{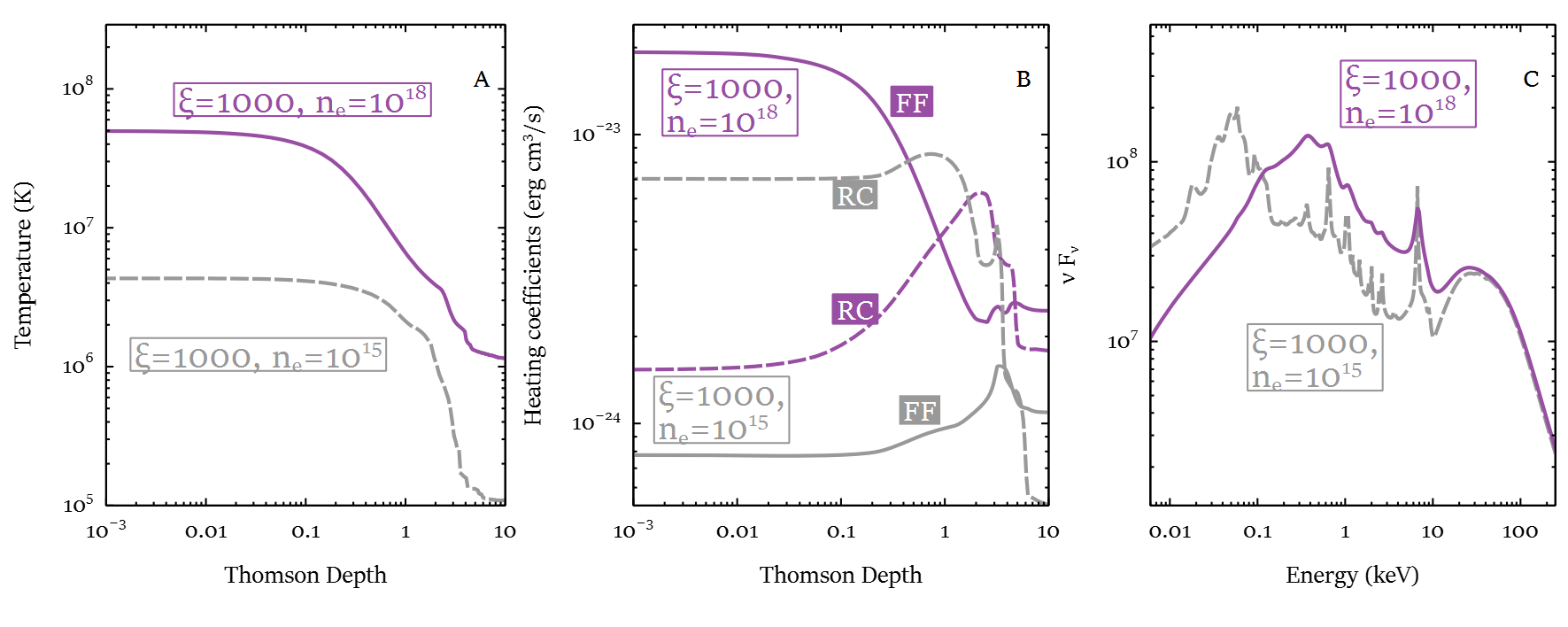}
    \caption{The temperature profiles of a disk with different electron densities are shown in the left panel calculated using the codes \texttt{Reflionx} \citep{ross07}. At~higher densities, the~disk surface is significantly hotter due to higher free-free (FF) absorption heating coefficients shown by the solid curves in the middle panel. The~dashed curves in the middle panel represent the heating coefficients due to the recombination cascade (RC). In~the higher temperature regime (e.g., $\tau < 1$), dielectronic recombination dominates over radiative and three-body recombination. The~dielectronic recombination rate decreases with increasing density. In~the lower temperature regime (e.g., $\tau > 1$), three-body recombination becomes dominant and increases with density. As~the density increases, the~dominant heating mechanism shifts from RC to FF, resulting in blackbody-like emission in the spectrum, as~shown in the right~panel.}
    \label{pic_reflionx}
\end{figure}

To further demonstrate the effects of high-density plasma, Figure\,\ref{pic_reflionx} shows the impact of high density on plasma emission, calculated using the \texttt{Reflionx} code \citep{ross07}. The~left panel shows that the disk surface temperature is significantly higher when the electron density is high. This is due to higher heating coefficients (heating rates divided by $n_{\rm e}n_{\rm H}$), as~in the middle panel. The~solid curves represent the heating coefficients due to free-free absorption (FF), which increase significantly with density, while the dashed curves show the heating coefficients due to recombination cascade (RC). Recombination processes in \texttt{Reflionx} have three main components: radiative recombination, three-body recombination, and~dielectronic recombination. Radiative recombination occurs when a positively charged ion captures an electron into one of its bound orbits, with~simultaneous emission of a photon. In~the dielectronic recombination process, the~energy released during the capture is used to promote a bound electron to another bound~orbit. 

At low temperatures, such as deeper into the disk where $\tau>1$, three-body recombination becomes significant at high densities compared to low densities. An~increase in density leads to a rise in the total recombination rate (compare the purple and grey dashed curves in the middle panel). At~higher temperatures, such as closer to the disk surface where $\tau<1$, dielectronic recombination dominates over radiative and three-body recombination. In~this regime, the~total recombination rate decreases as the density increases. 
This physical effect, first noted by \citet{burgess69}, was included in the ion balances of\linebreak \citet{summers74} and implemented in \texttt{Reflionx} by \citet{ross07}.

Overall, the~resulting spectra are shown in the right panel of Figure\,\ref{pic_reflionx}. The~ionisation and density choices in this figure approximate the best-fit parameters for the intermediate state observation of \src\ \citep{tomsick18}. A~high-density model exhibits blackbody-like emission in the soft X-ray band due to a higher temperature on the disk surface. Such features have been used to infer disk density from the data in previous studies (e.g., \cite{tomsick18,jiang19c}). There have also been recent developments in high-density models \citep{ding24} and attempts in fitting such models to additional objects (e.g., \cite{jiang22l,connors22,coughenour23}). In~the context of \src, such work still needs further exploration, including investigations into different flux states and the interplay between the disk and the~corona.

\unskip

\section{Conclusions} 

In this review, I have summarised some of the most important discoveries related to the BH HMXB \src. First, \src\ stands out as a unique example of a stellar-mass BH. Its young age suggests that the high BH spin inferred from X-ray data is more likely related to its stellar evolution history rather than its accretion history. Moreover, \src\ differs significantly in terms of mass and BH spin compared to BHs observed through GW detections, implying that they may represent distinct populations of~BHs.

\src\ is also an ideal laboratory for studying accretion physics. Numerous X-ray observations have revealed the presence of an accretion disk that extends close to the innermost stable circular orbit, consistent with the standard thin disk model. However, some open questions remain, such as the origin of the higher-than-expected polarisation degree measured by \textit{{IXPE}} and the dramatically different polarization angles observed in the hard X-ray to soft gamma-ray band compared to the radio jet orientation by \textit{{Integral}} and \textit{{AstroSAT}}. A~synergistic effort combining spectral, polarimetric, and~timing data is required, and~a comprehensive analysis using a consistent model has yet to be completed for \src.

Additionally, X-ray observations of \src\ provide an excellent opportunity to study high-energy plasmas. The~detection of hard X-ray emission above 511 keV points to the presence of non-thermal particle distributions, which has not been fully appreciated in current X-ray modelling. This is likely due to the absence of high-energy data, but~future missions like \textit{{COSI}} will be crucial in addressing this~gap.

In the soft X-ray band, \src\ offers the chance to test the standard thin disk model by comparing reflected emission spectra with photoionisation models, based on the expected plasma conditions in the disk. In~the intermediate state, the~inferred disk density for \src\ is around a thousand times higher than previously assumed. How this density changes with the spectral state and mass accretion rate remains an open question for future~studies.



\vspace{6pt}

\funding{J.J. acknowledges support from the Leverhulme Trust, Isaac Newton Trust and St Edmund's~College. J.J. would like to thank the useful discussion with John Tomsick and James Steiner.}

\conflictsofinterest{{  The funders had no role in the design of the study; in the collection, analyses, or interpretation of data; in the writing of the manuscript; or in the decision to publish the results. }
} 

\begin{adjustwidth}{-\extralength}{0cm}
\printendnotes[custom]

\reftitle{References}

\PublishersNote{}
\end{adjustwidth}
\end{document}